\documentclass[]{JHEP}

\usepackage{epsfig}

\def\bea{\begin{eqnarray}}
\def\eea{\end{eqnarray}}
\def\be{\begin{equation}}
\def\ee{\end{equation}}
\def\nn{\nonumber}
\def\ba{\begin{array}{l}}
\def\ea{\end{array}}

\newcommand\tw[2]{
 \Bigg[\hspace{-1pt}\raisebox{1pt}
 {$\begin{array}{c}
 \displaystyle{#1} \\ \displaystyle{#2}
 \end{array}$}
 \hspace{-1pt}\Bigg]}
\newcommand\pertw[4]{
 \Bigg[\hspace{-1pt}\raisebox{1pt}
 {$\begin{array}{c}
 \displaystyle{#1} \\ \displaystyle{#2}
 \end{array}$} 
 \hspace{0pt}\Bigg|\hspace{0pt}\raisebox{1pt}
 {$\begin{array}{c}
 \displaystyle{#3} \\ \displaystyle{#4}
 \end{array}$}
 \hspace{-1pt}\Bigg]}
\def\Z{{\bf Z}}

\title{On string models with Scherk--Schwarz 
supersymmetry breaking}

\author{Claudio A. Scrucca \\
CERN, 1211 Geneva 23, Switzerland \\
{\footnotesize \tt Claudio.Scrucca@cern.ch}}
\author{Marco Serone \\
ISAS-SISSA, Via Beirut 2-4, 34013 Trieste, Italy \\
INFN, sez. di Trieste, Italy \\
{\footnotesize \tt serone@sissa.it}}

\abstract{We construct a general class of chiral four-dimensional 
string models with Scherk--Schwarz supersymmetry breaking, involving 
freely acting orbifolds. The basic ingredient is to combine an ordinary 
supersymmetry-preserving $\Z_N$ projection with a supersymmetry-breaking 
projection $\Z_M^\prime$ acting freely on a subspace of the internal 
manifold. A crucial condition is that any generator of the full orbifold 
group $\Z_N \times \Z_M^\prime$ must either preserve some supersymmetry or 
act freely in order to become irrelevant in some large volume limit. 
Tachyons are found to be absent or limited to a given region of the 
tree-level moduli space. We find several new models with orthogonal 
supersymmetries preserved at distinct fixed-points. Particular attention 
is devoted to an interesting $\Z_3\times \Z_3^\prime$ heterotic example.}

\preprint{CERN-TH/2001-183 \\ SISSA-55/2001/EP \\ 
{\tt hep-th/0107159}}

\begin{document}

\section{Introduction}

It is commonly accepted that weak-scale Supersymmetry (SUSY) represents
a reasonable intermediate solution to the physics beyond the Standard Model
(SM). However, the mechanism of SUSY breaking is still a very open issue, 
in particular when trying to embed the SM or its Minimal Supersymmetric 
version (MSSM), in a fundamental theory including gravity, such as string 
theory. Ultimately, we are therefore still very far from any viable example 
of fundamental theory for particle physics. The traditional picture
for the latter is a string model with fundamental and compactification 
scales $M_{\rm s}$ and $M_{\rm c}$ of the order of the Planck scale, but it 
has now been understood that much lower $M_{\rm s}$ and/or $M_{\rm c}$ 
can actually be achieved \cite{extradim,milli,add}, making a low SUSY-breaking 
scale $M_{\rm SUSY}$ more natural.

One of the most interesting mechanisms of SUSY breaking is the so-called 
Scherk--Schwarz (SS) mechanism \cite{ss,css}, in which SUSY is broken at 
$M_{\rm c}$ by twisting the boundary conditions of each field through a 
global R-symmetry transformation. More in general, the same idea can be 
used to break also gauge symmetries, by supplementing the twist with a 
gauge transformation \cite{ssgauge}. This non-local breaking mechanism is 
very natural in the presence of compact dimensions, since the possibility 
of twisting boundary conditions is not forbidden by any symmetry of the 
theory; it can thus be considered as spontaneous, in this somewhat loose 
sense. Moreover, it is completely perturbative, and can therefore be 
efficiently investigated. 
These properties are quite appealing, especially from the string theory 
point of view, where it has been known for some time that the underlying 
superconformal structure forbids any continuous perturbative SUSY 
breaking \cite{stringtheorem}; the SS mechanism evades this theorem because 
SUSY is recovered only in a singular decompactification limit. 
As shown in \cite{extradim, Ito}, one more interesting property arises 
in the string context: only massless states give a sizeable one-loop 
contribution to the cosmological constant. More precisely, 
$\Lambda \sim (n_B-n_F)\,M_{\rm c}^4 + O(e^{-M_{\rm c}^2/M_{\rm s}^2})$, 
where $n_B$ and $n_F$ denote the number of massless 
bosons and fermions in the model. Unfortunately, this still yields an 
unacceptably large value, unless $n_B=n_F$. 

An important question to address for string models with SS SUSY breaking
is whether $M_{\rm susy}$ can be low enough, since it is set by the 
compactification scale $M_{\rm c}$. In oriented models, like heterotic 
orbifolds, $M_{\rm c}$ is naturally tied to the string scale 
$M_{\rm s}$, and both must be very close to the Planck scale in order to achieve the 
correct value of the Newton constant and perturbative gauge couplings in $D=4$.
Gauge coupling unification is then achieved around $M_s$.
On the contrary, this is instead generally lost by introducing a 
large hierarchy between $M_c$ and $M_s$. Moreover, large 
threshold corrections to gauge couplings \cite{thresh} arise in general
(see \cite{kkpr} for studies in models with partially 
broken SUSY), and much effort has been devoted in the past to finding models 
exempt of such corrections \cite{extradim,amq}.
New interesting possibilities arise instead for unoriented models, where 
$M_{\rm c}$ and $M_{\rm s}$ are less constrained and can be independently 
low.

A general method to construct string models with SS-type SUSY 
breaking by deforming supersymmetric orbifold models \cite{orb} has been 
developed in the past \cite{ssstringa}. It has been realized that basic 
principles of string theory, like modular invariance, pose severe 
restrictions on the implementation of the SS mechanism with this method, 
allowing in practice only discrete R-parity twists. 
Moreover, the range of application of this 
method appears to be limited to rather peculiar orbifolds \cite{b}, 
and unfortunately does not apply to the most interesting $\Z_3$ models
(see \cite{dg} for a field theory analysis). 
Subsequently, it has been realized that similar models could be 
obtained as freely acting\footnote{By freely acting, we always mean
an action that is free at least on a submanifold of the compactification 
manifold.} orbifolds \cite{freely}, opening in principle the 
possibility to construct a much larger class of models with SS SUSY 
breaking; however, little progress has been accomplished in 
this respect. Unoriented models can be obtained as orientifold descendants 
of these oriented models, and several examples have been worked out by now
\cite{ads,aads}.

Recently, renewed interest for the SS mechanism has emerged through a 
series of interesting papers considering five-dimensional orbifold 
field theories in which SUSY is broken at $M_{\rm c}$ to yield the 
SM at lower energies (see for instance \cite{phen,bhn}). 
A particularly simple and interesting example of this kind has been 
obtained in \cite{bhn}, by compactification on an orbifold of the type 
$S^1/\Z_2 \times \Z_2^\prime$, where the $\Z_2$ and $\Z_2^\prime$ actions 
preserve orthogonal supersymmetries and have fixed-points separated by 
a translation in $S^1$.
This construction can be reinterpreted as a SS compactification on the 
orbifold $S^1/\Z_2$, where a $\Z_2^\prime$ subgroup of the R-symmetry 
group is used to twist the boundary conditions, or similarly with $\Z_2$ and 
$\Z_2^\prime$ interchanged \cite{bhn}. It represents the simplest 
realization of SS SUSY breaking through a freely acting orbifold, 
and an embedding of this simple construction into a realistic string 
model is a very important challenge for the future.

The aim of this paper is to investigate in some generality the possibility
of constructing orbifold string models with SUSY broken \`a la SS, in 
which orthogonal supersymmetries are preserved at distinct fixed-points 
separated by a translation in the compact space. 
The main idea is to combine a standard SUSY-preserving action 
$G$ with a SUSY-breaking action $G^\prime$ acting freely in a 
subspace of the internal compactification torus, in such a way that the 
generators of the full $G \times G^\prime$ action consist of 
SUSY-preserving elements with disjoint sets of fixed-points, 
and freely acting SUSY-breaking elements mapping the fixed-points of 
each set into each other. Possible tachyons can arise only in twisted sectors 
of the freely acting elements, and are therefore massive over
most of the compactification moduli space.
We find that the allowed geometries are basically
either of the known $\Z_2 \times \Z_2^\prime$ type, possibly with an 
additional $\Z_K$ projection, or of a new $\Z_3 \times \Z_3^\prime$ type, 
and we construct a general class of examples of this kind of 
$\Z_N \times \Z_N^\prime$ models. 

As an interesting application, we present a novel class of four-dimensional
$\Z_3 \times \Z_3^\prime$ heterotic models with SS SUSY breaking.
For generic embedding of the orbifold action in the gauge bundle, one finds 
a chiral spectrum with a non-vanishing $n_B - n_F$ which can be either
positive or negative, but not zero. Also the number $n_T$ of would-be 
tachyons is in general non-zero, but there are a few examples with $n_T = 0$,
for which tachyons are therefore completely absent.

The paper is organized as follows. In section 2, we describe 
the main features of freely acting orbifolds that are relevant to our 
construction. In section 3, the partition functions of such models are 
derived and in section 4 their stability is briefly analysed.
In section 5 we present an explicit $\Z_N\times \Z_N^\prime$ construction, 
discuss their realization for $N=2,3$, and in section 6 we consider in 
more detail a $\Z_3\times \Z_3^\prime$ example.
In the last section we report our conclusions. Some modular properties of
orbifold partition functions are collected in an appendix.

\section{Freely acting orbifolds and SUSY breaking}

The type of orbifold models we are looking for can be described in very 
simple terms. Consider an orbifold group $G$ generated by a set of 
elements $\{g_i,g_j^\prime\}$ such that each $g_i$ acts non-freely, with 
fixed-points $P^k_{g_i}$, and preserves some SUSY, whereas the 
$g_j^\prime$ act freely, and do not preserve any SUSY. Clearly,
such a construction is highly constrained from the requirements of a finite 
group structure and modular invariance. Moreover, 
the freely acting generators $g_j^\prime$ should map a fixed-point $P^k_{g_i}$
of any of the non-freely acting element $g_i$ into another fixed-point 
$P^{k^\prime}_{g_i}$ of the same element $g_i$; this condition ensures, in particular,
that the orbifold is abelian. 

The crucial property characterizing SUSY breaking in such a model 
is the fact that the associated elements act freely. More precisely, they
must act as a simple translation by a finite fraction of lattice vector
in at least one of the 3 internal tori. This implies indeed that such 
elements trivialize in a suitable decompactification limit, in which 
SUSY is therefore restored. This is a clear implementation of 
the SS SUSY breaking mechanism in string theory and, interestingly enough, 
the same mechanism can be applied also to gauge symmetries by 
embedding non-trivially the $g_j^\prime$'s in the gauge bundle. 
Intuitively, it is obvious that thanks to the translation that they 
contain, the elements $g_j^\prime$ result effectively in the implementation 
of a twist around a given cycle of the internal space. Would the 
$g_j^\prime$'s act non-freely, then SUSY would be broken at the 
string scale rather than the compactification scale.

In the following, we construct explicit examples of the above type by 
considering product groups $G = \Z_M \times \Z_N^\prime$, in which the 
$\Z_M$ factor is generated by a non-freely acting SUSY-preserving 
element $g$, and the factor $\Z_N^\prime$ by a freely acting SUSY-breaking 
element $g^\prime$. In order to obtain the required structure, one must
then analyse the action of each generator of the full $G$. To this aim,
it is convenient to recall at this stage some basic facts about 
supersymmetries in four-dimensional orbifold compactifications. 
The basic Majorana--Weyl supercharge $Q$ in $D=10$ fills the 
${\bf 16}$ of $SO(9,1)$.
This decomposes in $D=4$ into four Majorana supercharges 
$Q_n = {Q_n}_L + {Q_n}_R$, transforming each as a 
${\bf 2} \oplus {\bf \bar 2}$ under $SO(3,1)$ and together as a 
${\bf 4}$ of the maximal $SO(6)$ R-symmetry group. 
For each $n=1,2,3,4$, ${Q_n}_{L}$ and ${Q_n}_{R}$ have $SO(6)$ weights 
$w_n$ and $-w_n$ respectively, where:
$$
w_1 = \mbox{$(\frac 12,\frac 12,\frac 12)$} \;,\;\;
w_2 = \mbox{$(\frac 12,-\frac 12,-\frac 12)$} \;,\;\;
w_3 = \mbox{$(-\frac 12,\frac 12,-\frac 12)$} \;,\;\;
w_4 = \mbox{$(-\frac 12,-\frac 12,\frac 12)$} \;.
$$

A generic orbifold element $g$ acts as the combination 
of a rotation of angle $2 \pi v_i$ and some unspecified shift, in each 
of the 3 internal $T^2_i$. Under this action, the 4 possible supercharges 
transform as:
\bea
&&{Q_n}_L \rightarrow  e^{2 \pi i v \cdot w_n} {Q_n}_L \,, \nn \\
&&{Q_n}_R \rightarrow  e^{-2 \pi i v \cdot w_n} {Q_n}_R \,. \nn 
\eea
Therefore, the supercharge $Q_n$ is left invariant by $g$ if 
$v \cdot w_n$ is an integer, independently of the shift.

\section{Partition functions}

In order to construct explicit examples of the models described in
the previous section, the basic two-dimensional blocks of the partition 
function for a generic twist involving both a rotation and a translation 
are needed. These are used to derive the full heterotic and Type II
partition functions of such orbifolds and deduce the constraints 
arising from the requirement of modular invariance. 
Type I open descendants could be constructed in the
usual way, with the additional constraint of tadpole cancellation,
but we shall not consider such constructions here. Most of the results
reported below are standard \cite{mod1,mod2}, but we have re-analysed them 
without any assumption about SUSY, in order to avoid any possible
confusion.

Consider first a $\Z_N$ group generated by the element 
$\alpha = \alpha_{\rm geom} \alpha_{\rm gauge}$, where 
$\alpha_{\rm geom}$ defines the geometric action on the internal 
compactification torus and $\alpha_{\rm gauge}$ is its embedding in
the gauge bundle. Take the geometric part to be
\be
\alpha_{\rm geom} = \exp 2 \pi i \sum_{i=1}^3 \Big(v_i J^i 
+ R_i \delta_i P^i \Big) \,,
\label{alphageom}
\ee
with $J^i$ and $P^i$ being the generators of rotations and diagonal 
translations in each internal two-torus $T^2_i$ with basic radii $R_i$. 
The gauge part is of course trivial for Type IIB models, whereas for
the $E_8 \times E_8$ heterotic string, it has the general form
\be
\alpha_{\rm gauge} = \exp 2 \pi i \sum_{p=1}^8 
\Big(v^\prime_p J^{\prime p} + v^{\prime\prime}_q J^{\prime\prime q}\Big) \,,
\label{alphagauge}
\ee
with $J^{\prime p}$ and $J^{\prime\prime q}$ being the Cartan currents
of the two $E_8$ factors. 

In order to have $\alpha^N = 1$, one must take $v_i = r_i/N$, 
$v_p^\prime = r_p^\prime/N$ and $v_q^{\prime\prime} = r_q^{\prime\prime}/N$
with integer $r_i$, $r_p^\prime$ and $r_q^{\prime\prime}$, and due to
spinor representations, impose the constraints
\be
N \Big(\sum_{i=1}^3 v_i, \sum_{p=1}^8 v_p^\prime,
\sum_{q=1}^8 v_q^{\prime\prime} \Big) = 0 \;{\rm mod}\; 2 \,.
\label{spin}
\ee

The partition function for one complex field with twists 
$(g,h) = (k v,l v)$, shifts $(\hat g,\hat h) = (k \delta, l \delta)$, 
and spin structure $(a,b)$ ($k,l=0,1,...,N-1$; $a,b = 0,\frac 12$) is easily
computed. For a complete (left + right) boson, one finds:
\bea
&& Z_B \pertw{h}{g}{\hat h}{\hat g}(\tau) = \left\{
\begin{array}{ll}
|\eta(\tau)|^{-4} \Lambda \tw{\hat h}{\hat g}(\tau) 
&\;,\;\;\mbox{if $(g,h) = (0,0)$} \nn \\
\displaystyle{\Bigg|\eta(\tau) \, 
\theta^{-1}{\frac 12 + h \brack \frac 12 + g}(\tau)\Bigg|^2} 
&\;,\;\;\mbox{if $(g,h) \neq (0,0)$} \nn
\end{array}
\right. \;.
\eea
The lattice contribution is given by (see \cite{freely}):
\bea
\Lambda \tw{\hat h}{\hat g}(\tau) &=& 
\frac{\sqrt{G}}{\alpha^\prime\,{\rm Im}\,\tau} \,
\sum_{\vec m, \vec n} e^{- \frac {\pi }{\alpha^\prime{\rm Im}\,\tau} 
[(m + \hat g) + (n + \hat h)\tau]_i (G+B)_{ij} 
[(m + \hat g) + (n + \hat h) \bar \tau]_j} \nn \\
&=& \sum_{\vec m, \vec n} e^{2 \pi i \,\hat g \cdot m} \, 
q^{\frac 12 |P_L[\hat h]|^2} 
\,\bar q^{\frac 12 |P_R[\hat h]|^2} \;,
\label{lattice}
\eea
where $\sqrt{G}=\sqrt{{\rm det} G_{ij}}$ is related to the volume 
$V$ of the target-space torus $T^2$ by $V = (2 \pi)^2 \sqrt{G}$ and 
the lattice momenta are given by
\bea
P_L[\hat h] &=& \frac 1{\sqrt{2\,{\rm Im}\,T\,{\rm Im}\, U}} 
\Bigg[- m_1\, U + m_2 +  T\,\Big((n_1+\hat h) + (n_2+\hat h)\,U\Big)
\Bigg] \;, \nn \\
P_R[\hat h] &=& \frac 1{\sqrt{2\,{\rm Im}\,T\,{\rm Im}\, U}} 
\Bigg[- m_1\, U + m_2 + \bar T\, \Big((n_1+\hat h) + (n_2+\hat h)\,U\Big) 
\Bigg] \;,
\eea
in terms of the standard dimensionless moduli $T$ and $U$ parametrizing 
the metric $G$ and antisymmetric field $B$ as:
\be
G_{ij} = \alpha^\prime \frac {{\rm Im}\,T}{{\rm Im}\,U} \left(
\begin{array}{cc}
1 & {\rm Re}\,U \\
{\rm Re}\,U & |U|^2 
\end{array}
\right) \;,\;\;
B_{ij} = \alpha^\prime \left(
\begin{array}{cc}
0 & {\rm Re}\,T \\
- {\rm Re}\,T & 0 
\end{array}
\right) \;.
\ee
Notice that (\ref{lattice}) reduces to 
$V/(4 \pi^2 \alpha^\prime\,{\rm Im}\,\tau)$ for a non-compact boson. 

For a fermion, one finds instead
\bea
Z_F \pertw{a}{b}{h}{g}(\tau) &=& \eta^{-1}(\tau)\,e^{-2 \pi i b h}\,
\theta{a+h \brack b+g}(\tau) \nn \\
&=& \eta^{-1}(\tau)\, \sum_{p_a=n+a} 
q^{\frac 12 (p_a + h)^2} e^{2 \pi i (p_a + h) g} e^{2 \pi i p_a b} \,.
\label{ZF}
\eea
Notice in particular the crucial phase in the first row of
(\ref{ZF})\footnote{This phase is very often missing in the literature, 
probably because it is irrelevant for supersymmetric constructions.
It has previously been noticed in \cite{mod2}, but is not evident in 
\cite{mod1}.}, ensuring that the GSO projection amounts to the standard 
constraints on the bosonization momentum $p$ independently of $h$, as 
evident from the second row of (\ref{ZF}).

The modular properties of these basic partition functions, as well as the 
constraints they impose, 
are derived in the appendix; we report here only the main final results. 
Denoting the generic twist with
$G = (g_i,\hat g_i,g_i^\prime,g_i^{\prime\prime})$ and its conjugate 
with $\bar G = (N v_i - g_i, 1 - \hat g_i, N v_i^\prime - g_i^\prime,
N v_i^{\prime\prime} - g_i^{\prime\prime})$, the total partition 
function\footnote{To be precise, we consider the light-cone partition
function; the contribution of longitudinal degrees of freedom provides
only the correct invariant measure $d^2\tau/{\rm Im}\,\tau^2$ in the 
world-sheet moduli space.} is given by
\be
Z = \sum_{G,H} C \tw{H}{G} N \tw{H}{G} Z \tw{H}{G} \,,
\label{Ztotale}
\ee
where $Z {H \brack G}$, $N {H \brack G}$ and $C {H \brack G}$ denote
respectively the partition function, the number of fixed-points and
an arbitrary overall phase in each sector ${H \brack G}$.

For a generic Type IIB model, one finds
\bea
Z \tw{H}{G} &=& Z_B \pertw{0}{0}{0}{0} 
\prod_{i=1}^3 Z_B \pertw{h_i}{g_i}{\hat h_i}{\hat g_i}
\Bigg|\sum_{a,b=0,\frac 12} \!\eta_{ab} \, 
Z_F \pertw{a}{b}{0}{0} \prod_{i=1}^3 Z_F \pertw{a}{b}{h_i}{g_i}\Bigg|^2 
\,, \label{ZtotaleIIB} \\
C \tw{H}{G} &=& 1 \,,
\label{CIIB}
\eea
where $\eta_{ab} = (-1)^{2a + 2b + 4ab}$ are the usual GSO-projection signs.
The partition function is then modular-invariant without any further 
restriction.

For a generic heterotic $E_8\times E_8$ model one gets:
\bea
Z \tw{H}{G} &=& Z_B \pertw{0}{0}{0}{0} 
\prod_{i=1}^3 Z_B \pertw{h_i}{g_i}{\hat h_i}{\hat g_i}
\Bigg(\frac 12 \sum_{a,b=0,\frac 12} \!\eta_{ab} \, 
Z_F \pertw{a}{b}{0}{0} \prod_{i=1}^3 Z_F \pertw{a}{b}{h_i}{g_i}\Bigg) \nn \\
&\;& \times \, \Bigg(\frac 12 \sum_{c,d=0,\frac 12} \prod_{p=1}^8
\bar Z_F \pertw{c}{d}{h_p^\prime}{g_p^\prime} \times 
\frac 12 \sum_{e,f=0,\frac 12} \prod_{q=1}^8 
\bar Z_F \pertw{e}{f}{h_q^{\prime\prime}}{g_q^{\prime\prime}}\Bigg) 
\,, \label{Ztotalehet} \\
C \tw {H}{G} &=& e^{- i \pi (g \cdot h - g^\prime \cdot h^\prime 
- g^{\prime\prime} \cdot h^{\prime\prime})} \,.
\label{Chet}
\eea
This partition function is modular-invariant provided that the embedding
satisfy the condition \cite{mod1,mod2} (see also \cite{imnq}):
\be
N (v^2 - v^{\prime 2} - v^{\prime\prime 2}) = 0 \:{\rm mod}\; 2 \,.
\label{modinv}
\ee

It is straightforward to extend this analysis to more general orbifold 
groups $G$ with more than one generator. The results 
(\ref{ZtotaleIIB})-(\ref{Chet}) hold true, but the condition 
(\ref{modinv}) must be extended to all the independent generators of $G$.

Recall finally that modular invariance of the partition function, together 
with the condition of tadpole cancellation for unoriented descendants, 
guarantees the full consistency of this kind of string models, and implies 
in particular the absence of divergences or anomalies. Actually, as shown 
in \cite{hetano,ssunor}, even the complete mechanism of anomaly cancellation 
is encoded in the background dependence of the partition function itself
in a very natural way.

\section{Stability}

An important issue for the kind of non-supersymmetric models we aim to 
construct is their stability. At tree level, one must make sure that no 
tachyonic modes appear. 

To check the presence of tachyons, one must compute the zero-point energy.
The contribution of each left or right complex field can be easily read 
off from the behaviour of the corresponding partition function in the limit 
${\rm Im}\,\tau \rightarrow \infty$. One finds:
\bea
&& E^0_B [h] =  \frac 1{24} 
- \frac 12 \Big(\frac 12 - \theta[\mbox{$\frac 12$}|h]\Big)^2 \,, \\
&& E^0_F [a|h] =  -\frac 1{24} + \frac 12 \Big(a - \theta[a|h]\Big)^2 \,,
\eea
where
\be
\theta[a|h] = |h| - \mbox{int}(|h| + \mbox{$\frac 12$} - a) \,.
\ee
In the following, it will prove convenient to use the bosonized 
description for all the fermions. Correspondingly, one must count a 
zero-point energy of only $E^0_F[0|0] = -1/24$ for each of them, since 
in this description the orbifold action is a lattice shift and no longer
a twist. It is then useful to define the quantity
\be
C[h] = \frac 12 \sum_{i=1}^3 
\theta[\mbox{$\frac 12$}|h_i]\Big(1-\theta[\mbox{$\frac 12$}|h_i]\Big) \,.
\ee
For Type IIB models, the mass formula is:
\bea
&& L_0[a|h] =
\frac 12 |P_L[\hat h]|^2
+  N_L[h] + \frac 12 (p_a + h)^2 + \Big(- \frac 12 + C[h] \Big) \,, \nn \\
&& \bar L_0[c|h] =
\frac 12 |P_R[\hat h]|^2
+  N_R[h] + \frac 12 (p_c + h)^2 + \Big(- \frac 12 + C[h] \Big) \,. \nn
\eea
For heterotic models, one finds similarly:
\bea
&& L_0[a|h] = 
\frac 12 |P_L[\hat h]|^2
+  N_L[h] + \frac 12 (p_a + h)^2 + \Big(-\frac 12 + C[h]\Big) \,, \nn \\
&& \bar L_0[c|e|h] = 
\frac 12 |P_R[\hat h]|^2
+  N_R[h] + \frac 12 (p_c^\prime + h^\prime)^2 
+ \frac 12 (p_e^{\prime\prime} + h^{\prime\prime})^2 
+ \Big(- 1 + C[h]\Big) \,. \nn
\eea
The sectors $a,c,e$ are now associated to the different classes of lattice
vectors. 

For each state, the Hamiltonian $H[h] = L_0[h] + \bar L_0[h]$ gives the mass 
squared as $m^2[h] = \frac 2{\alpha^\prime} H[h]$, whereas the level mismatch
$\Delta[h] = L_0[h] - \bar L_0[h]$ is related by modular invariance 
to the phase picked under the orbifold transformation, which reads 
$\phi(g) = e^{2 \pi i \Delta[g]}$ for an orbifold transformation 
$g$\footnote{Notice for instance that the contribution to $\Delta[g]$ 
from KK and winding modes is given by 
$\frac 12 (|P_L[\hat g]|^2 - |P_R[\hat g]|^2) = m \!\cdot\! (n+\hat g)$ 
and leads to the phase $e^{2 \pi i m \cdot \hat g}$, in agreement 
with (\ref{lattice}).}. The level-matching condition 
$L_0[h] = \bar L_0[h]$ therefore implies invariance under 
the orbifold action in twisted sectors.

Tachyons can occur only in twisted sectors associated to the SUSY-breaking 
elements. The mass of a generic state in these sectors is of the form 
$m^2 = m_0^2 + \frac 1{\alpha^\prime}(|P_L|^2 + |P_R|^2)$,
with a moduli-independent contribution $m_0^2$, which can be either
positive or negative, but a positive-definite moduli-dependent contribution
from the internal momentum. It is then quite clear that possible tachyons 
can always be made massive by selecting a suitable part of moduli space. 
More precisely, the minimal value of $|P_L|^2 + |P_R|^2$ is 
generically obtained for the zero mode $m=n=0$; one then finds:
\be
m^2 > m_0^2 + \frac {{\hat h}^2}{\alpha^\prime} 
\frac {|T(1+U)|^2}{{\rm Im\,T}\,{\rm Im\,U}} \,.
\ee
Focusing for concreteness on a fixed complex structure $U$, the condition 
for the absence of tachyons turns into a restriction on the K\"ahler 
modulus $T$. One finds:
\be
({\rm Re}\,T)^2 + ({\rm Im}\,T - T_0)^2 > T_0^2 \,,
\ee
with
\be
T_0 = \frac {\alpha^\prime(-m_0^2)}{2\,{\hat h}^2} 
\frac {{\rm Im\,U}}{|1+U|^2} \,.
\label{T0}
\ee
This condition excludes a circle close to the origin in the moduli space
of $T = (B + i\,R^2)/\alpha^\prime$, as depicted in fig. 1. 
Notice in particular that only $R \geq \sqrt{T_0\,\alpha^\prime}$ is allowed 
for $B=0$, but all the $R > 0$ are allowed as soon as 
$B \geq T_0\,\alpha^\prime$.

\begin{figure}[h]
\begin{picture}(300,190)(0,0)
\put(110,10){\epsfig{file=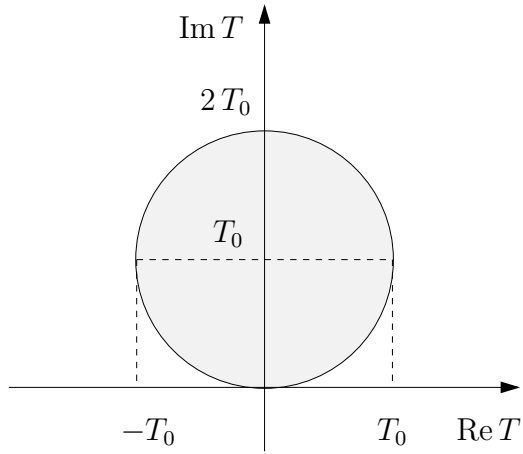,height=6cm}}
\put(175,167){${\rm Im}\,T$}
\put(188,90){$T_0$}
\put(183,140){$2\,T_0$}
\put(153,15){$-T_0$}
\put(250,15){$T_0$}
\put(280,15){${\rm Re}\,T$}
\end{picture}
\caption{Tachyons can arise only in the shaded region of the $T$ moduli 
space.}
\end{figure}

It is clear from the above discussion that the stability of this kind of 
model is triggered by the effective potential of both the radion field and 
its pseudoscalar partner, entering together in a would-be chiral multiplet 
$T$ of the low-energy effective action.
This potential is completely flat at the tree level, but since SUSY
is broken, non-trivial quantum corrections are expected to occur, and the 
VEV of $T$ is dynamically fixed. 

\section{Explicit constructions}

The general construction described so far actually admits relatively few 
concrete realizations. We restrict ourselves to 
abelian orbifolds involving rotations and translations in the lattice of 
the internal $T^6$. The basic group structure is of the form 
$\Z_M \times \Z_N^\prime$, where $\Z_M$ is generated by a standard 
SUSY-preserving non-freely acting rotation $\alpha$, and $\Z_N^\prime$ 
by the combination $\beta$ of a SUSY-breaking rotation and a translation. 
We focus in the following on the case $M=N$.

The partition function of such $\Z_N(\alpha) \times \Z_N^\prime(\beta)$ 
models has the form:
\be
Z = \frac 1{N^2} \sum_{k,l=0}^{N-1} \sum_{p,q=0}^{N-1} 
N \tw{\alpha^l\beta^{q}}{\alpha^k\beta^{p}}
Z \tw{\alpha^l\beta^{q}}{\alpha^k\beta^{p}} \,.
\ee
Interestingly, this partition function can be decomposed into $\Z_N$ blocks.
To see this, let us define $\alpha_i = \alpha \beta^{i-1}$, $i=1,\cdots,N$, 
so that the total orbifold group can be rewritten as 
$G = \{1,\alpha_i, \cdots, \alpha_i^{N-1}, \beta, \cdots, \beta^{N-1}\}$.
The new elements $\alpha_i$ are combinations of SUSY-preserving rotations 
and translations, with fixed-points $P^a_{\alpha_i}$ differing from one 
another by the translation contained in $\beta$ (see figs. 2 and 3). 
In this new parametrization, the partition function simplifies 
substantially: only those sectors of the form 
${\alpha_i^q \brack \alpha_i^p}$ or ${\beta^q \brack \beta^p}$ give
a non-vanishing contribution to the partition function. In this sense,
one can therefore write: $\Z_N(\alpha) \times \Z_N^\prime(\beta) = 
\Z_N(\alpha_1) + \cdots + \Z_N(\alpha_{N-1}) + \Z_N(\beta)$, and the
partition function can be rewritten as:
\be
Z = Z^{U}_{\Z_N(\alpha_i) \times \Z_N(\beta)} 
+ \sum_{i=1}^N \frac {N_{\Z_N(\alpha_i)}}N Z^{T}_{\Z_N(\alpha_i)}
+ \frac {N_{\Z_N(\beta)}}N Z^{T}_{\Z_N(\beta)} \,,
\ee
where
\bea
&& Z^{U}_{\Z_N(\alpha_i) \times \Z_N(\beta)} = 
\frac 1{N^2} \Bigg[Z \tw {0}{0} + \sum_{b=1}^{N-1} 
\Bigg(\sum_{i=1}^{N} Z \tw {0}{\alpha_i^b} + Z \tw {0}{\beta^b} \Bigg)
\Bigg] \,, \hspace{25pt} \\
&& Z^{T}_{\Z_N(\alpha_i)} = \frac 1N 
\sum_{a=1}^{N-1}\sum_{b=0}^{N-1} Z \tw {\alpha_i^a}{\alpha_i^b} \,, \\
&& Z^{T}_{\Z_N(\beta)} = \frac 1N
\sum_{a=1}^{N-1}\sum_{b=0}^{N-1} Z \tw {\beta^a}{\beta^b} \,.
\eea
The untwisted sector can be computed by projecting the usual untwisted 
sector of any of the supersymmetric $\Z_N(\alpha_i)$ orbifolds with the 
SUSY-breaking action $\Z_N(\beta)$. The choice of $\alpha_i$ is irrelevant, 
since $\Z_N(\alpha_i) \times \Z_N(\beta) = G$ for any $\alpha_i$. The massless 
states of the $\Z_N(\alpha_i)$ orbifold that are not invariant under 
$\Z_N(\beta)$ will survive only as KK or winding modes, and get a mass of 
order $M_{c}$. There are then the twisted sectors of the supersymmetric 
$\Z_N(\alpha_i)$ orbifolds and those of the non-superymmetric
$\Z_N(\beta)$ orbifold, all with a degeneracy given by the number
of fixed-points divided by a factor $N$.
This additional factor has a clear geometric interpretation. 
For $\alpha_i$ twisted sectors, it reflects 
the fact that not all the fixed-points are independent; they fall into groups 
of $N$ filling orbits of $\beta$. For $\beta$ twisted sectors, all the 
fixed-hyperplanes are independent, but they fill the $T^2$ where the shift 
acts, and there is therefore an additional factor of $1/N$ from the volume.

In these models, SUSY is broken at a scale set by the volume 
of the $T^2$ where $\beta$ acts as a shift: $M_{\rm SUSY} = M_{\rm c}$. 
This is due to the shift entering the definition of $\beta$. 
Indeed, this has two crucial consequences. The first is that the 
SUSY-breaking element $\beta$ trivializes for large $R$. 
The second is that the $N$ elements $\alpha_i$, preserving 
different fractions of the maximal SUSY, have fixed-points which 
differ by a fraction of lattice vectors, and therefore move far 
apart for large $R$.
Potential tachyons can appear only in the non-supersymmetric $\Z_N(\beta)$ 
twisted sectors. As explained in section 4, all the states in these sectors 
have a moduli-dependent positive contribution to their mass squared,
and tachyons can thus always be avoided.

Out of these basic $\Z_N \times \Z_N^\prime$ models, one can then 
in general construct more complicated models by further projecting 
with an additional $\Z_K$ action generated by a SUSY-preserving rotation 
$\gamma$, which is orthogonal to the translation in $\beta$. We report 
in the following the examples that we have been able to construct.

\subsection{$\Z_2 \times \Z_2^\prime$ models}

Consider the orbifold group $G = \Z_2 \times \Z_2^\prime$,
where the two factors are generated by the elements
\bea
&& \alpha: v_\alpha = \mbox{$(\frac 12,\frac 12,0)$} \;,\;\; 
\delta_\alpha = \mbox{$(0,0,0)$} \,; \\
&& \beta: v_\beta = \mbox{$(0,1,0)$} \;,\;\; 
\delta_\beta = \mbox{$(\frac 12,0,0)$} \,.
\eea
This kind of models have already been considered in \cite{freely,ads}.
Defining $\alpha_i = \alpha \beta^{i-1}$, the orbifold group 
can be rewritten as $G = \{1,\alpha_i, \beta\}$, where
\be
\begin{array}{ll}
\alpha_1 : &\mbox{preserves $Q_2$ and $Q_3$} \,; \nn \\
\alpha_2 : &\mbox{preserves $Q_1$ and $Q_4$} \,; \nn \\
\beta    : &\mbox{does not preserve any $Q_n$} \,. \nn
\end{array}
\label{22}
\ee

\begin{figure}[h]
\begin{picture}(300,210)(0,0)
\put(100,10){\epsfig{file=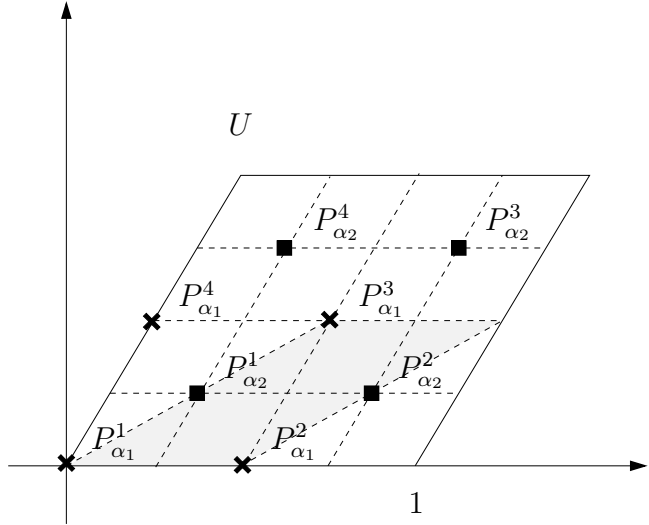,height=7cm}}
\put(184,158){$U$}
\put(252,15){$1$}
\put(132,39){$P_{\alpha_1}^1$}
\put(200,39){$P_{\alpha_1}^2$}
\put(165,93){$P_{\alpha_1}^4$}
\put(233,93){$P_{\alpha_1}^3$}
\put(182,66){$P_{\alpha_2}^1$}
\put(248,66){$P_{\alpha_2}^2$}
\put(216,122){$P_{\alpha_2}^4$}
\put(281,122){$P_{\alpha_2}^3$}

\end{picture}
\caption{The $\alpha_i$ fixed-points $P_{\alpha_i}^a$ in $T^2_1$ for the 
$\Z_2\times \Z_2^\prime$ model.
The SUSY-breaking element $\beta$ acts as a shift in this plane,
and relates different fixed-points of the same element $\alpha_i$,
$\beta: P_{\alpha_i}^a \rightarrow P_{\alpha_i}^{a+2}$. One can take
$P_{\alpha_i}^{1,2}$ as independent fixed-points. Correspondingly, the 
fundamental cell of the orbifold theory can be chosen to be the shaded 
area, since this is mapped to the fundamental cell of the whole torus
through $\alpha_i$ and $\beta$ transformations.}
\end{figure}

This model can be lifted to $D=6$, where it represents the unique possibility 
of a model with $N=1 \rightarrow N=0$ SUSY breaking. From
the $D=4$ point of view, however, it has $N=2 \rightarrow N=0$ SUSY 
breaking and is therefore non-chiral. More interesting $D=4$ models 
with orbifold group $G=\Z_2 \times \Z_K \times \Z_2^\prime$ can be 
obtained by a further $\Z_K$ orbifold projection acting in the last 
two $T^2$'s, which does not influence the freely acting SUSY-breaking 
element. One can choose the generator $\gamma$ of this action to have 
$v_\gamma = (0,\frac 1K,\frac 1K)$ and $\delta_\gamma = (0,0,0)$.
It is then straightforward to show that all the elements in $G$ either 
preserve some SUSY or act freely in the first $T^2$. More precisely,
for $k=1,...,K-1$, one finds that, in addition to the conditions (\ref{22}),
$\gamma^n$ preserves $Q_3$ and $Q_4$, $\alpha \gamma^k$ preserves at 
least $Q_3$, 
$\alpha \gamma^k \beta$ preserves at least $Q_4$, whereas 
$\gamma^n \beta$ do not preserve any $Q_n$ in general
but act as translations in the first $T^2$. The resulting models have 
therefore all $N=1 \rightarrow N=0$ SUSY breaking. The cases 
$K=2,3$ have already been discussed in \cite{aads}.

\subsection{$\Z_3 \times \Z_3^\prime$ models}

Consider now the orbifold group $G = \Z_3 \times \Z_3^\prime$,
where the two factors are generated by the elements:
\bea
&& \alpha: v_\alpha = \mbox{$(\frac 13,\frac 13,0)$} \;,\;\; 
\delta_\alpha = \mbox{$(0,0,0)$} \,; \\
&& \beta: v_\beta = \mbox{$(0,0,\frac 23)$} \;,\;\; 
\delta_\beta = \mbox{$(\frac 13,0,0)$} \,.
\eea
Different choices for the SUSY-preserving element $\alpha$ lead to 
equivalent models, and no other options are possible for the SUSY-breaking 
element $\beta$, so that this construction is essentially unique.
For instance, an equivalent model would have been obtained by considering
the usual $N=1$ supersymmetric $\Z_3$ twist for $v_\alpha$.
Defining $\alpha_i = \alpha \beta^{i-1}$, the total orbifold group can 
be rewritten as $G = \{1, \alpha_i, \alpha_i^2, \beta,\beta^2\}$, where:
\be
\begin{array}{ll}
\alpha_1 : &\mbox{preserves $Q_2$ and $Q_3$} \,; \nn \\
\alpha_2 : &\mbox{preserves $Q_4$} \,; \nn \\
\alpha_3 : &\mbox{preserves $Q_1$} \,; \nn \\
\beta: &\mbox{does not preserve any $Q_n$} \,. \nn
\end{array}
\ee
Notice that this class of models does not have $N=2$ twisted sectors along
the SUSY-breaking directions, implying that most likely no threshold 
corrections will depend on the corresponding moduli.

\begin{figure}[h]
\begin{picture}(300,220)(0,0)
\put(90,10){\epsfig{file=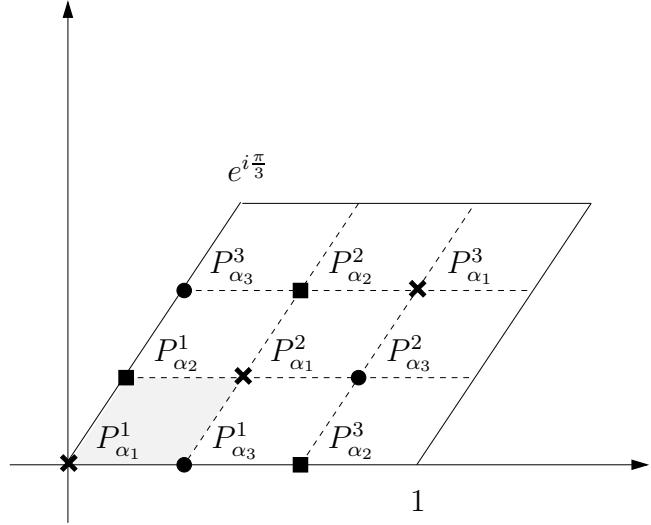,height=7cm}}
\put(173,140){$e^{i \frac \pi 3}$}
\put(242,15){$1$}
\put(123,39){$P_{\alpha_1}^1$}
\put(167,39){$P_{\alpha_3}^1$}
\put(211,39){$P_{\alpha_2}^3$}
\put(145,72){$P_{\alpha_2}^1$}
\put(189,72){$P_{\alpha_1}^2$}
\put(233,72){$P_{\alpha_3}^2$}
\put(166,105){$P_{\alpha_3}^3$}
\put(211,105){$P_{\alpha_2}^2$}
\put(256,105){$P_{\alpha_1}^3$}
\end{picture}
\caption{The $\alpha_i$ fixed-points $P_{\alpha_i}^a$ in $T^2_1$ for the 
$\Z_3\times \Z_3^\prime$ model.
Again, $\beta$ acts as a shift in this plane, and relates different 
fixed-points of the same $\alpha_i$, 
$\beta: P_{\alpha_i}^a \rightarrow P_{\alpha_i}^{a+1}$. One can take
$P_{\alpha_i}^1$ as independent fixed-points and the shaded area as 
the fundamental cell of the orbifold theory.}
\end{figure}

\section{The $\Z_3\times \Z_3^\prime$ heterotic model}

Consider the $\Z_3 \times \Z_3^\prime$ construction introduced in the previous 
section applied to heterotic strings. The condition (\ref{modinv}), evaluated 
for all the independent generators, implies that the embeddings should satisfy:
\be
v_\alpha^{\prime 2} + v_\alpha^{\prime\prime 2} 
= \frac 29 \;\mbox{mod}\; \frac 23 \;,\;\;
v_\beta^{\prime 2} + v_\beta^{\prime\prime 2}
= \frac 49 \;\mbox{mod}\; \frac 23 \;,\;\;
v_\alpha^{\prime} v_\beta^{\prime} 
+ v_\alpha^{\prime\prime} v_\beta^{\prime\prime}
= 0 \;\mbox{mod}\; \frac 23 \;.
\label{emb33}
\ee
We will analyse in detail the case of standard embedding of both 
actions into the gauge bundle as an example, and then discuss qualitative 
features of more general embeddings.

\subsection{Standard embedding}

Consider first the untwisted sector. This is most easily derived 
starting from the $N=2$ $\Z_3(\alpha)$ model, and further projecting 
the spectrum by $\Z_3(\beta)$. Massless left-moving states are associated 
with lattice 4-vectors $p$ with $p^2 = 1$ and 
$\sum_m p_m = \mbox{odd}$, filling the ${\bf 8_V}$ 
and ${\bf 8_S}$
of $SO(8)$. These states pick up a phase 
$\phi_{\alpha}(p) = e^{2 \pi i p \cdot v_{\alpha}}$ 
under $\alpha$ transformations, and a phase 
$\phi_{\beta}(p) = e^{2 \pi i p \cdot v_{\beta}}$ 
under $\beta$ transformations. 
Denoting with $\alpha = e^{\frac {2\pi}3 i}$ and 
$\beta = e^{\frac {2\pi}3 i}$ the basic phases under these two 
transformations, one finds the following decomposition:
\bea
{\bf 8_V} &\rightarrow& \Big[{\bf 2_V}\Big] \,\oplus\, 
\Big[2\cdot{\bf 1}\Big] (\alpha + \alpha^{-1}) \,\oplus\, 
\Big[{\bf 1}\Big](\beta + \beta^{-1}) \,, \nn \\ 
{\bf 8_S} &\rightarrow& \Big[2\cdot{\bf 1}\Big](\beta + \beta^{-1}) \,\oplus\, 
\Big[{\bf 1}\Big](\alpha + \alpha^{-1})(\beta + \beta^{-1}) \,. 
\eea
There are then several relevant types of right-moving states.
Neutral states arise from right-moving states with 
$p=p^\prime=p^{\prime\prime} = 0$ and $N_R = 1$, and fill 
an ${\bf 8_V}$ of $SO(8)$. Under the orbifold action, they decompose 
as their left-mover counterparts. Charged states under each $E_8$ factor 
are instead associated to right-moving lattice 8-vectors $p^\prime$ or 
$p^{\prime\prime}$ with $p^{\prime 2},p^{\prime\prime 2} = 0,2$ and 
$\sum_m p^\prime_m = \mbox{even}$, corresponding to $N_R = 1,0$.
These fill a ${\bf 120}$ and a ${\bf 128}$ 
of $SO(16)$, forming in total the ${\bf 248}$ of $E_8$.
The hidden sector is unaffected by the orbifold projection. 
In the visible sector, the $\Z_3(\alpha)$ projection breaks $E_8$ to 
$E_7 \times U(1)$, whereas the $\Z_3(\beta)$ projection further breaks 
this to $SO(10) \times SU(2) \times U(1)^2$ and makes all the gauginos 
massive; charged states pick up a phase $\phi_{\alpha}(p^\prime) = 
e^{-2 \pi i p^\prime \cdot v_{\alpha}}$
under $\alpha$ transformations, and a phase $\phi_{\beta}(p) = 
e^{-2 \pi i p^\prime \cdot v_{\beta}}$
under $\beta$ transformations,
and decompose as follows\footnote{In the following, 
for simplicity we will not distinguish between ${\bf 16}$ and 
${\bf\overline{16}}$, ${\bf 2}$ and ${\bf\overline 2}$ representations. 
For the same reason, we will not report the $U(1)$ charges of the states. 
One can easily check that the spectrum is chiral.}:
\bea
{\bf 120} &\rightarrow& 
\Big[({\bf 45},{\bf 1}) \oplus ({\bf 1},{\bf 3}) 
\oplus 2\,({\bf 1},{\bf 1})\Big] \,\oplus\,
\Big[({\bf 10},{\bf 2}) \oplus ({\bf 1},{\bf 1})\Big] 
(\alpha + \alpha^{-1}) \nn \\ &\;& \,\oplus\,
\Big[({\bf 10},{\bf 1})\Big] (\beta + \beta^{-1}) \,\oplus\, 
\Big[({\bf 1},{\bf 2})\Big] (\alpha + \alpha^{-1}) (\beta + \beta^{-1}) \;,
\nn \\ 
{\bf 128} &\rightarrow& 
\Big[({\bf 16},{\bf 2})\Big] (\beta + \beta^{-1}) \,\oplus\,
\Big[({\bf 16},{\bf 1})\Big] (\alpha + \alpha^{-1}) (\beta + \beta^{-1}) \;.
\eea
The massless spectrum is found by tensoring the above left and 
right-moving states and keeping only invariant states. In this
way, one finds a total content in the untwisted sector which can be 
summarized as follows:
\bea
{\bf 8_V} \otimes {\bf 8_V} &:& {\bf 2_V} \otimes {\bf 2_V} \,\oplus\, 
10 \;; \nn \\
{\bf 8_S} \otimes {\bf 8_V} &:& 4 \;; \nn \\
{\bf 8_V} \otimes {\bf 248} &:& {\bf 2_V} \otimes 
\Big[({\bf 45},{\bf 1}) \oplus ({\bf 1},{\bf 3}) \oplus 
2\,({\bf 1},{\bf 1})\Big] \nn \\
&\;& \,\oplus\, 4\,({\bf 10},{\bf 2}) \,\oplus\, 2\,({\bf 10},{\bf 1}) 
\,\oplus\, 4\,({\bf 1},{\bf 1}) \,\oplus\, 2\,({\bf 16},{\bf 2}) \nn \;; \\
{\bf 8_S} \otimes {\bf 248} &:& 4\,({\bf 10},{\bf 1}) 
\,\oplus\, 4\,({\bf 1},{\bf 2}) \,\oplus\, 4\,({\bf 16},{\bf 2}) 
\,\oplus\, 4\,({\bf 16},{\bf 1}) \;; \nn \\
{\bf 8_V} \otimes {\bf 248^\prime} &:& {\bf 2_V} \otimes {\bf 248^\prime} 
\;; \nn \\
{\bf 8_S} \otimes {\bf 248^\prime} &:& - \; .
\eea

Consider next $\alpha_i$-twisted sectors.
The $\alpha_1$-twisted sector preserves $N=2$ SUSY, and the spectrum 
of hypermultiplets is known: at each of the $9$ fixed-planes one gets 
1 ${\bf 56}$ and 7 singlets of $E_7$. Each of the $\alpha_{2,3}$-twisted 
sectors preserves instead a $N=1$ SUSY, and the spectrum of chiral 
multiplets is similar for both of them: at each of the $27$ fixed-points 
one gets a $({\bf 27},{\bf 1})$ and 3 copies of $({\bf 1},{\bf \bar 3})$ 
of $E_6 \times SU(3)$. Decomposing into representations of 
$SO(10) \times SU(2)$, one finds in total:
\bea
\alpha_1 &:& \mbox{$3$ hyper-mult. in }\Big[
2\,({\bf 16},{\bf 1}) \,\oplus\, 
({\bf 10},{\bf 2}) \,\oplus\,
2\,({\bf 1},{\bf 2}) \,\oplus\,
7\,({\bf 1},{\bf 1})
\Big] \nn \;; \\
\alpha_2 &:& \mbox{$9$ chiral-mult. in }\Big[
({\bf 16},{\bf 1}) \,\oplus\, 
({\bf 10},{\bf 1}) \,\oplus\,
3\,({\bf 1},{\bf 2}) \,\oplus\,
4\,({\bf 1},{\bf 1})
\Big] \nn \;; \\
\alpha_3 &:& \mbox{$9$ chiral-mult. in }\Big[
({\bf 16},{\bf 1}) \,\oplus\, 
({\bf 10},{\bf 1}) \,\oplus\,
3\,({\bf 1},{\bf 2}) \,\oplus\,
4\,({\bf 1},{\bf 1})
\Big] \nn \;.
\eea

Finally, consider the $\beta$-twisted sectors, where potential tachyonic
states might arise. In the left-moving sector, there is only one such state 
in the NS sector with $(p+v)^2 = \frac 19$ and $N_L=0$. 
For right-movers, there are $14$ such states with 
$(p^\prime+v^\prime)^2 = \frac {10}9$ and $N_R=0$.
Pairing these states, one finds would-be tachyons in the 
$({\bf 10},{\bf 1}) \oplus 2\,({\bf 1},{\bf 2})$ of
$SO(10) \times SU(2)$ with $\frac {\alpha^\prime}2 m^2 = - \frac 23 + 
\frac 12 (|P_L|^2[\hat h] + |P_R|^2[\hat h])$. The level-matching 
condition selects the KK and winding modes satisfying 
$m(n+h) = \mbox{integer}$, allowing $m=0$ mod $3$. The worst situation 
arises for $m=n=0$, and using $m_0^2 = -\frac 4{3\alpha^\prime}$ and 
$U = e^{i\frac \pi 3}$, one computes from (\ref{T0}) that $T_0 = \sqrt{3}$. 
For $|T - i\,T_0| > T_0$, all the states in these sectors are 
massive\footnote{Notice, however, that since these tachyons are in the 
appropriate representation to correspond to the SM Higgs, one may wish to 
keep some of them and impose a less restrictive constraint.}.

Summarizing, the model we have constructed exhibits a chiral spectrum,
and SUSY is broken at the scale $M_{\rm SUSY} = R_1^{-1}$ 
together with part of the gauge group. 
Tachyons can be avoided independently of $R_1^{-1}$ by choosing 
$B_1 > \sqrt{3}\,\alpha^\prime$. One has $n_B - n_F = 534$ from the 
untwisted sector. There are then 3 supersymmetric twisted sectors, clearly 
with $n_B - n_F = 0$. Finally, there is one non-supersymmetric twisted sector,
which gives $n_B - n_F = 0$ if $|T - i\,T_0| > T_0$ and $n_B - n_F = 14$ if 
$|T - i\,T_0| = T_0$.

\subsection{More general embeddings}

Models with more general embeddings can be easily constructed,
and there are actually only a few possibilities to explore. As usual, 
thanks to the symmetries of the $E_8$ lattice, one can restrict shift 
vectors with length squared less than 1, whose embeddings must 
satisfy the conditions (\ref{emb33}). Moreover, when applied 3 times, 
they must reduce to a lattice vector; since the lattice is even, this 
implies that $v_\alpha^{\prime 2}, v_\alpha^{\prime\prime 2},
v_\beta^{\prime 2}, v_\beta^{\prime\prime 2} = 0 \;\mbox{mod}\; \frac 49$.
The first condition in (\ref{emb33}) leaves then 9 independent 
possibilities for $(v_\alpha^{\prime 2}, v_\alpha^{\prime\prime 2})$, namely:
$(0, \frac 29)$, $(\frac 29 ,0)$, $(0,\frac 89)$, $(\frac 89, 0)$, 
$(\frac 29,\frac 23)$, $(\frac 23, \frac 29)$, $(\frac 49,\frac 49)$, 
$(\frac 23, \frac 89)$, $(\frac 89, \frac 23)$. 
Similarly, the second condition in (\ref{emb33}) restricts
$(v_\beta^{\prime 2}, v_\beta^{\prime\prime 2})$ to be among the following 
8 possibilities: $(\frac 29,\frac 29)$, $(0,\frac 49)$, $(\frac 49,0)$, 
$(\frac 23, \frac 49)$, $(\frac 49,\frac 23)$, $(\frac 29, \frac 89)$, 
$(\frac 89,\frac 29)$, $(\frac 89,\frac 89)$. 
As in \cite{orb}, one can then choose a unique representative $w$ for each
value of $w^2$.
Finally, the last condition in (\ref{emb33}) turns out to constrain only 
the relative permutations of the shift vectors for the two factors. 

We have computed the mismatch $n_B-n_F$ between massless bosons and 
fermions, as well as the number $n_T$ of would-be tachyons, for all these 
models. The possible values for $(n_B-n_F,n_T)$ depend only on the embedding 
$(v_\beta^{\prime 2},v_\beta^{\prime\prime 2})$ of the SUSY-breaking element
$\beta$, and one finds 
$(318,2)$ for $(\frac 29,\frac 29)$,
$(534,14)$ for $(0,\frac 49)$ or $(\frac 49,0)$, 
$(48,2)$ for $(\frac 23,\frac 49)$ or $(\frac 49,\frac 23)$,
$(156,8)$ for $(\frac 29,\frac 89)$ or $(\frac 89,\frac 29)$,
and $(-6,0)$ for $(\frac 89,\frac 89)$.
There exist therefore models without any possible tachyons\footnote{In these cases, 
one could define consistent models with hard 
SUSY breaking at the string scale by dropping the shift in the 
SUSY-breaking element $\beta$.}. Notice also that for generic embeddings,
it becomes particularly clear that the class of models under consideration 
is intrinsically chiral. Indeed, each sector, and in particular the two 
$N=1$ sectors, will have in general a different gauge twist, leading to 
distinct spectra of representations.

We did not consider the additional freedom of adding Wilson lines in
our models. It should be appreciated, however, that like most of the 
other moduli, Wilson lines are now dynamical. A non-trivial 
effective potential is generated for these 
gauge-invariant operators \cite{hos}, that will in general dynamically break 
part of the gauge group.

\section{Conclusions}

In this paper, a new class of four-dimensional non-supersymmetric string 
vacua has been analysed. The key point of the construction resides in the 
idea of considering freely acting translations and
non-supersymmetric rotations, in addition to standard supersymmetric 
orbifold rotations.
In this way, SUSY is broken at the compactification scale $M_{\rm c}$
through a string version of the Scherk--Schwarz mechanism \cite{freely}. 
Although we focused on simple examples of oriented closed string models, 
our construction is quite general and can be easily generalized in various 
way. For example, one can construct more complicated models with different 
supersymmetries or gauge symmetries being broken at different 
compactification scales $M_{\rm c}^i$. Unoriented models with D-branes and 
O-planes can be derived from Type IIB models as in \cite{ads,aads}. 

There are, we believe, several interesting issues that deserve further study.
Among all, the most important would be a deeper analysis of the quantum 
stability of these models. In particular, one should study the quantum 
effective potential for the compactification moduli to see whether a 
stabilization of the geometry can be achieved. A similar question should be 
faced also for the dilaton and for Wilson lines, whose VEV's are also 
dynamically determined at the quantum level. 
Using the by now well established string web of dualities, it would also 
be instructive to analyse possible dual realizations of our models,
allowing to study their strong coupling behaviour.

Finally, we think that the new SUSY-breaking geometries found in this paper 
are quite promising from the point of view of realistic model building.
As already noticed in the introduction, it would be exciting to embed in
this kind of string models the recently constructed higher-dimensional 
field theory models with SS SUSY and gauge symmetry breaking 
\cite{phen,gauge}. The possibility of having an exponentially suppressed 
cosmological constant \cite{Ito,extradim} is also quite appealing in this
context.

\vskip 14pt
\noindent
{\large \bf Acknowledgements}
\vskip 7pt

\noindent
We would like to thank C. Angelantonj, I. Antoniadis, A. Sagnotti and 
A. Uranga for valuable discussions and useful comments. This work has 
been partially supported by the EEC through the RTN network 
``The quantum structure of space-time and the geometric nature of 
fundamental interactions'', contract HPRN-CT-2000-00131."
     
\appendix

\section{$\theta$-functions and modular invariance}

Defining $q = \exp 2 \pi i \tau$, one has
\be
\eta(\tau) = q^{\frac 1{24}} \prod_{n=1}^\infty (1 - q^n) \,,
\ee
and 
\bea
\theta \tw {a}{b}(\tau)
&=& \sum_n q^{\frac 12(n+a)^2} e^{2 \pi i (n+a)b} \nn \\
&=& e^{2 \pi i a b} q^{\frac {a^2}2}
\prod_{n=1}^\infty (1 - q^n)
(1 + q^{n+a-\frac 12} e^{2 \pi i b})
(1 + q^{n-a-\frac 12} e^{-2 \pi i b}) \,,
\eea
satisfying the periodicity property
\be
\theta \tw {a+m}{b+n}(\tau) = e^{2 \pi i n a} \theta \tw {a}{b}(\tau) \,.
\ee
Under modular transformations, these functions transform as follows:
\bea
&& \eta(\tau + 1) = e^{i \frac \pi{12}} \,\eta(\tau) \,, \\
&& \eta(-1/\tau) = \sqrt{-i\tau} \,\eta(\tau) \,,
\eea
and
\bea
&& \theta \tw {a}{b}(\tau + 1) = e^{-i\pi a(a-1)}\,
\theta \tw {a}{a + b - \mbox{$\frac 12$}}(\tau) \,, \\
&& \theta \tw {a}{b}(- 1/\tau) = \sqrt{-i\tau}\,
e^{2 \pi i ab} \, \theta \tw {b}{-a}(\tau) \,.
\eea

Using the above formulae, the modular properties of the basic partition 
functions reported in section 3 are the following:
\bea
&& Z_B \pertw{h}{g}{\hat h}{\hat g}(\tau + 1) = 
Z_B \pertw{h}{g+h}{\hat h}{\hat g + \hat h}(\tau) \,, \\
&& Z_B \pertw{h}{g}{\hat h}{\hat g}(-1/\tau) = 
Z_B \pertw{g}{N v - h}{\hat g}{1-\hat h}(\tau) \,,
\eea
and 
\bea
&& Z_F \pertw{a}{b}{h}{g}(\tau + 1) = 
e^{- i \frac \pi{12} - i \pi [a(a-1) + h^2]} 
Z_F \pertw{a}{a+b-\mbox{$\frac 12$}}{h}{g+h}(\tau) \,, \\
&& Z_F \pertw{a}{b}{h}{g}(-1/\tau) = 
e^{- 2 \pi i [ab + g (N v - h) + b N v]} 
Z_F \pertw{b}{a}{g}{N v - h}(\tau) \,.
\eea

At this point, it is straightforward to derive the conditions required
to achieve modular invariance of the general partition function 
(\ref{Ztotale}). A basic modular transformation maps the ${H \brack G}$ 
sector into either the ${H \brack G+H}$ or ${G \brack \bar H}$ sectors, 
and since
\be
N \tw{H}{G+H} = N \tw{G}{\bar H} = N \tw{H}{G} \,,
\ee
the residual phases occurring in this transformation will severely constrain
the $C {H \brack G}$'s.

For a generic Type IIB model, the partition function (\ref{ZtotaleIIB})
in a generic sector is found to transform (using (\ref{spin})) as:
\bea
&& Z \tw{H}{G}(\tau + 1) = 
Z \tw{H}{G+H}(\tau) \,, \\
&& Z \tw{H}{G}(-1/\tau) = 
Z \tw{G}{\bar H}(\tau) \,.
\eea
One can therefore take $C {H \brack G} = 1$ as in (\ref{CIIB}), 
without any further condition.

For a generic heterotic $E_8 \times E_8$ model, again using (\ref{spin}),
the partition function (\ref{Ztotalehet}) transforms as: 
\bea
&& Z \tw{H}{G}(\tau + 1) = 
e^{- i \pi (h^2 - h^{\prime 2} - h^{\prime\prime 2})}
Z \tw{H}{G+H}(\tau) \,, \\
&& Z \tw{H}{G}(-1/\tau) = 
e^{- 2 \pi i [g \cdot (N v - h) - g^\prime \cdot (N v^\prime - h^\prime)
- g^{\prime\prime} \cdot (N v^{\prime\prime} - h^{\prime\prime})]}
Z \tw{G}{\bar H}(\tau) \,.
\eea
These transformations leave the partition function invariant if the 
$C {H \brack G}$'s satisfy
\bea
&&C \tw {H}{G+H} = 
e^{- i \pi (h^2 - h^{\prime 2} - h^{\prime\prime 2})} \,C \tw {H}{G} 
\label{condc1} \,, \\
&&C \tw {G}{\bar H} = 
e^{- 2 \pi i [g \cdot (N v - h) - g^\prime \cdot (N v^\prime - h^\prime)
- g^{\prime\prime} \cdot (N v^{\prime\prime} - h^{\prime\prime})]}\, 
C \tw {H}{G} \label{condc2} \,.
\eea
An additional consistency condition arises in this case from the requirement
that each sector ${H \brack G}$ should be separately invariant under 
$\tau \rightarrow \tau + N$. This happens if
$N (v^2 - v^{\prime 2} - v^{\prime\prime 2}) = 0 \:{\rm mod}\; 2$,
as anticipated in (\ref{modinv}). In this case, all the phases proportional 
to $N$ drop from (\ref{condc2}), and the conditions 
(\ref{condc1}) and (\ref{condc2}) have the unique solution 
$ C {H \brack G} = e^{- i \pi (g \cdot h - g^\prime \cdot h^\prime 
- g^{\prime\prime} \cdot h^{\prime\prime})}$, reported in (\ref{Chet}).
This factor is identified with the total phase picked up by the 
vacuum $|\Omega[H]\rangle$ in the $H$-twisted sector under the 
orbifold action $G$. This can be verified explicitly by constructing
this vacuum with twist fields. Notice that for standard embedding 
($v^\prime = v$, $v^{\prime\prime} = 0$) this is equal to $1$ in all 
sectors.

\end{document}